\documentclass[11pt,twoside]{article}
\usepackage[normalem]{ulem}
\usepackage[T1]{fontenc}
\usepackage[utf8]{inputenc}
\usepackage{newtxtext,newtxmath}
\usepackage{fancyhdr}
\usepackage{longtable}
\usepackage{geometry}
\usepackage{xpatch}
\usepackage{amsmath,amsfonts}
\usepackage{graphicx}
\usepackage{booktabs}
\usepackage{array}
\usepackage{tabularx}
\usepackage{hyperref}
\usepackage[style=apa,backend=biber]{biblatex} 
\DeclareFieldFormat[article]{title}{\mkbibemph{#1}}
\renewbibmacro{in:}{}

\hypersetup{
  colorlinks=true,
  linkcolor=blue,
  citecolor=blue,
  urlcolor=blue
}

\DeclareFieldFormat[article]{title}{\mkbibemph{#1}}
\DeclareFieldFormat[book]{title}{\mkbibemph{#1}}
\DeclareFieldFormat[inbook]{title}{\mkbibemph{#1}}
\DeclareFieldFormat[thesis]{title}{\mkbibemph{#1}}
\DeclareFieldFormat{journaltitle}{«#1»,}
\renewbibmacro{in:}{}
\DeclareFieldFormat{volume}{#1}
\DeclareFieldFormat{pages}{pp. #1}

\xpatchbibmacro{date+extrayear}{%
  \printtext[parens]%
}{%
  \setunit{\addspace}%
  \printtext%
}{}{}

\makeatletter

\def\@maketitle{%
     \newpage
     \null
     \vskip 2em%
     {\large \@author \par}%
     \vskip 1em%
     {\LARGE \normalfont \@title \par}%
     \vskip 2em%
}

\renewcommand\section{\@startsection{section}{1}{\z@}%
     {-3.5ex \@plus -1ex \@minus -.2ex}%
     {2.3ex \@plus.2ex}%
     {\normalfont\large\itshape}}

\renewcommand\subsection{\@startsection{subsection}{2}{\z@}%
     {-3.25ex\@plus -1ex \@minus -.2ex}%
     {1.5ex \@plus .2ex}%
     {\normalfont\normalsize\itshape}}

\renewcommand\subsubsection{\@startsection{subsubsection}{3}{\z@}%
     {-3.25ex\@plus -1ex \@minus -.2ex}%
     {1.5ex \@plus .2ex}%
     {\normalfont\normalsize\itshape}}

\renewenvironment{abstract}{%
     \small
     \noindent{\itshape Abstract}\par
     \vspace{0.5em}
}{}

\makeatother
\newcommand{\rev}[1]{#1}
\newcommand{\prf}[1]{#1}

\title{Finance professionals' demands on AI: beyond accuracy and explainability}
\author{Patricia Marcella Evite, Ekaterina Svetlova and Doina Bucur}

\begin{document}

\maketitle

\begin{abstract}
Recent scholarship on explainable artificial intelligence or XAI \prf{has} focused on the tradeoffs between accuracy of machine learning models and the explainability of its decisions. This negative relationship is not consistently observed and is highly context-dependent. The focus on this tradeoff also reflects more the needs of AI developers than finance professionals, whose preferences in such competing objectives in AI deployment we know little about. To address this gap, the authors examine how these so-called business users prioritize explainability—operationalized here as ease of understanding for non-technical users—relative to accuracy and other domain-specific considerations such as compliance, cost, and speed. Using twenty semi-structured interviews across banks, consultancies, and finance-serving AI vendors, we ask participants to rank the five factors and analyze the reasons for their choices. Findings revealed \prf{that} accuracy and compliance are perceived as the most important, although these patterns are not broadly generalizable because of the Europe-skewed sample. Results also show that preferences vary substantially even among participants within the same bank, underscoring the audience-dependent nature of AI deployment needs. As an exploratory study, the paper extends prior research that has typically considered only three factors in model selection by grounding a broader framework in practitioners’ perspectives.
\end{abstract}

\vspace{0.4cm}

{\small \noindent\textit{Keywords}\par
explainable artificial intelligence, finance, machine learning

}

\section{Introduction}

Explainable AI (XAI) is an emerging field that aims to address the opacity of highly performing models used in a wide array of financial decision-making--from credit scoring to green portfolio labeling-- through techniques that provide explanations for decisions made by AI \parencite{gohel2021}. A central theme in XAI research is the trade-off between explainability and accuracy  \parencite{herm2023}, \rev{also called the Performance-Explainability Tradeoff (PET) \parencite{crook2023revisiting}.} Essentially, the more accurate the AI model is, the harder it is to understand what constitutes the model’s decision. Yet, this trade-off is not always linear and is highly context-dependent  \parencite{goethals2022}. \rev{\textcite{crook2023revisiting} \prf{suggest} that this supposed negative relationship must be analyzed in a nuanced way, considering resource availability, domain characteristics, and risk.} For instance, a professional in healthcare may prioritize accuracy, while in criminal justice, explainability and accuracy may be valued equally \parencite{vanderveer2021}.

\rev{In 2025, a U.S. trading firm said they have ``surrendered to the machines''  and would worry about what the AI did only on the bad days \parencite{ft_machines_2025}, thus choosing profit-maximizing performance over explainability. However, there are heterogeneous goals in finance,} rendering this binary view insufficient. Financial AI products operate in a complex landscape shaped by trade-offs between predictive performance, regulatory compliance, and ethical requirements \rev{ \parencite{saxena2026ethical}. For instance, regulatory institutions involved in the classic loan application example would prioritize the capability of XAI tools to comply \prf{with} the law like the E.U. General Data Protection Regulation (GDPR) \parencite{state2025explanation}. 

Moreover, the accuracy-explainability issue reflects more the priorities of developers that involve model-specific concerns, differing from that of} \textit{business users}, those responsible for interacting with, interpreting, and defending AI-mediated decisions within financial institutions \parencite{arrieta2020}. \rev{Non-technical users are not only concerned with explainability, but also risks, privacy, and broader ethical deployment of AI systems  \parencite{amarnath2024revolutionizing, ioannou2026role}. Without these considerations, we may overlook the real impact of AI in organizations, its people, and the material environment it operates in \parencite{glaser2021}, particularly in finance \prf{where} decisions affect people and markets far beyond a single firm \parencite{Svetlova2022AI}.} 

\rev{\textcite{bahlool2026performance} notes that aspects like fairness are considered in isolation from the accuracy-explainability tradeoffs in AI-based credit scoring.  
But like most of the extant literature, researchers \prf{have}expressed constraints like compliance \parencite{saxena2026ethical}, privacy \parencite{shokri2021privacy,saifullah2024privacy, abbas2024imfblog}, or temporal budget \parencite{crook2023revisiting} within a multi-objective optimization problem and therefore, do not inform about how they are considered in operational situations by business users. The other strand of the available work involves frameworks (c.f. \textcite{yuvarani2025fairness}) or is normative in nature as to how to reconcile the tradeoffs with user expectations (c.f.  \textcite{crook2023revisiting}).  \textcite{bell2022s} argues that tradeoff is rarely characterized empirically.}

As such, \rev{we complement previous work by approaching the gap in an empirical manner similar to \textcite{liedeker2024empirical} who investigated the tradeoff between explanation complexity and value, and inquire in the finance field.} Accordingly, we introduce in this exploratory paper our research question:\\

\textbf{\textit{How do different business users in finance prioritize explainability relative to other competing factors in practice?}}\\\\
\rev{We have three contributions to the discourse. First, we extend the PET framing in XAI by examining explainability as one consideration among a broader set of operational demands in finance, namely compliance, cost, and speed, further expanding the three-factor model by \textcite{crook2023revisiting}. Second, drawing on interviews with finance professionals, our paper offers an empirically grounded account of how these actors prioritize and justify these competing demands in practice. Lastly, we illustrate how these tradeoffs also differ among various professionals within the same institutions, giving a more nuanced audience-dependent angle to the topic, since most studies are at the institutional level (c.f. \textcite{kuiper2021}).}

\rev{The remainder of this paper is structured as follows. Section \ref{tradeoff_lit} reviews the performance–explainability tradeoff and the XAI techniques used to manage it. Section \ref{understanding} operationalizes ease of understanding for non-technical audiences. Section \ref{tensions} enumerates competing tensions in the literature that go beyond the classical accuracy–explainability frontier. Section \ref{method} outlines the methodology, ethical considerations, and limitations. Section \ref{context} presents the empirical context, covering the AI use cases and organisational landscape of the participants. Section \ref{tradeoff_results} presents the findings on how practitioners prioritize the five factors, including audience-dependent tradeoffs within the same institution and other considerations raised beyond the five presented. Section \ref{conclusion} concludes and suggests avenues for further research.}

\rev{\section{The performance--explainability tradeoff}} \label{tradeoff_lit}
The next section reviews the performance--explainability tradeoff in financial AI and places the present study within that broader debate. It also introduces the explanation methods most often used to manage model opacity, showing why the concerns of business users extend beyond technical performance alone.

The models that deliver the most predictive accuracy \rev{(used interchangeably in the literature, and in this paper, with performance)}are typically the least transparent. \rev{For example, a linear regression model is usually easier to interpret because each coefficient shows how a variable affects the prediction, whereas a neural network can capture more complex patterns but is much harder to explain step by step. Other models like \prf{logic-based} models can be marginally behind neural networks but significantly outperform others in terms of explainability \parencite{adams2020type}. \textcite{wanjohi2024comparing} concluded a similar pattern with predictive power of multilayer perceptron algorithm outperforming the more interpretable logistic regression in credit risk.} This intensifies demands for explainability from supervisors (e.g., central banks), boards, and internal model risk teams \parencite{aquilina2025}.  \rev{In finance, the negative relationship is found in various studies from credit assessments to bitcoin forecasts \parencite{adams2020type, giudici2024explainable, thomas2026machine}, but not for other contexts since the inherently interpretable models achieved the best accuracy for corporate recovery bonds \parencite{nazemi2024interpretable} or  in fraud detection contexts \parencite{almalki2025financial}, to name a few. Even in the same goal like credit assessments, the results vary on the models compared and type of data (cf. \textcite{thomas2026machine} vs \textcite{obermann2016interpretable}).}

 \rev{PET is said to be taken for granted in AI studies, not just in finance, \parencite{crook2023revisiting} and has consequently been doubted in empirical settings \parencite{rudin2019stop, Gosiewska2021Simpler}}\rev{\textcite{petkovic2023not} also argues that performance and explainability should not be seen as a tradeoff in the first place because both are required for trustworthy AI.  In \textcite{bell2022s}'s pursuit for an empirical grounding of the binary tradeoff, they find no evidence for a consistent negative relationship, i.e., a \% increase in explainability would decrease performance by a certain \% of accuracy. Instead, the relationship is more nuanced in line with the proposition of \textcite{crook2023revisiting} and there were even cases where the tradeoff does not exist at all similar to the conclusions of \parencite{centofanti2022exploring}. }

To manage this opacity, financial institutions increasingly deploy XAI techniques. Comprehensive surveys distinguish between intrinsically interpretable \textit{(ante-hoc)} models and \textit{post-hoc} explanation methods layered on top of complex models. Ante-hoc models (e.g., linear scorecards, decision trees, rule lists) build interpretability directly into the model; they remain common in regulated credit risk because they are easy to audit but may sacrifice predictive performance on complex, high-dimensional data. Post-hoc model-agnostic explainers are now standard wherever non-linear ML models are used, notably, SHAP (SHapley Additive exPlanations),  LIME (Local Interpretable Model-agnostic Explanations), and counterfactual explanations \parencite{cerneviciene2024explainable, YeoVanDerHeeverMaoEtAl2025}.

More recently, institutions are experimenting with large language models(LLM)-based explanation layers on top of existing models (see for example \textcite{sivamayilvelan2025building} on deep learning for debt collection). Here, LLMs, or LLMs combined with Retrieval-Augmented Generation (RAG), are used to translate SHAP values, rule outputs, or policy logic into natural-language narratives for relationship managers, call-center staff, or compliance officers. Early work on RAG for financial documents suggests that grounding LLMs in bank reports and internal documents can improve factual accuracy and user trust when answering questions about financial statements or policies \parencite{iaroshev2024}. However, these LLM explanation layers introduce a new explainability concern: the risk that explanations are plausible but not faithful to the underlying model logic \parencite{jacovi2020, agarwal2024faithfulness}.

\section{Ease of understanding: operationalizing explainability for non-technical audiences} \label{understanding}
Here, we clarify how explainability is treated in this study. Rather than using it as a purely technical benchmark, we approach it as ease of understanding for the finance professionals who must work with, justify, or oversee AI-assisted decisions.

\rev{Although XAI enhances decision quality}, much of the classical XAI literature, particularly on SHAP, LIME, and counterfactuals, remains closer to computer-centric evaluation, where explanations are judged by how well they approximate the model or satisfy formal criteria such as faithfulness \parencite{lopes2022}, rather than by how well non-technical professionals can make sense of them in context. \rev{This has led to a crucial gap \prf{in} user-centric approach that considers non-technical \prf{users'} needs and understanding of explanations \parencite{naveed2022}, making current XAI methods unable to cater to preferences like concise and contextually visual explanations \parencite{Maathuis2026Designing}}. By contrast, our empirical focus is on business users and operators of financial AI systems rather than model developers \parencite{arrieta2020}. In line with  \textcite{tomsett2018}'s role-based model of interpretability, we concentrate on roles such as: operators and decision-ratifiers who must justify AI-assisted decisions to clients or internal committees; domain experts who validate model outputs against financial logic and institutional policy; risk and compliance professionals who must ensure that AI use remains within regulatory and ethical bounds. 

Audience-dependent studies in financial XAI show that these stakeholders are often educated and quantitatively literate but not AI specialists \parencite{jiang2022who}. They typically do not need to understand the full internal mechanics of gradient boosting or neural networks; instead, they need explanations that are usable for decision-making, documentation, and accountability, e.g., to write a file note to a supervisor or answer a regulator’s question \parencite{hadjimisheva2021}. Cognizant of this need, proposals have been developed to insert visual layers and feedback mechanisms directly into XAI models (see for example, \textcite{bucur2025}).

\rev{For our target participants}, we conceptualize the relevant property for our study as ``ease of understanding” of the model’s outputs and reasoning, rather than the \rev{computer-centric explainability since studies cite that the success of XAI in financial credit assessment depends on the balancing of complexity of explanations with user comprehension \parencite{zeng2022toward}. This is consistent with human-centered evaluation of XAI systems done by \textcite{Maathuis2026Designing}. Other empirical studies like \textcite{bell2022s} also include user understanding in explainability in addition to what the model does when they investigated tradeoffs, as opposed to how the model works.}

\rev{\section{Competing Tensions}} \label{tensions}
\rev{\textcite{Wilson2025CFA} enumerates three challenges in implementing XAI in finance, namely, technical challenges in benchmarks universalization and real-time decision making constraints; regulatory challenges in privacy and explainability standards; and lastly, user experience issues where there is lack of accessible interfaces for business users, regulators, and customers. But, rarely is the case that these issues are considered holistically in research \parencite{bahlool2026performance}. In fraud detection models, for example, \textcite{nesvijevskaia2021accuracy} discussed how performance is not the only consideration since banks are constrained by legal, operational, strategic factors. In this section, we enumerate some of the research that have moved beyond the classical accuracy--explainability frontier by treating explainability as one objective among several competing requirements.}

\rev{\subsection{Satisfaction and Value}
\textcite{liedeker2024empirical} studied whether there is a tradeoff between explanation complexity and value. Value encompassed five dimensions including perceived accuracy, understandability, plausibility, sufficiency of detail, and satisfaction. They \prf{found} that there are explanations that receive best evaluation in all dimensions and those are characterized by medium complexity, which they see as a \textit{sweet spot.}

We acknowledge that user satisfaction is top priority in human-centric XAI, but for the purpose of tradeoffs analysis, we exclude \prf{it} because it can mean a multitude of values within (c.f. \textcite{Hoffman2019Metrics}).}

\rev{\subsection{Compliance}
There are many standards in deployment of AI, most notably, the E.U. GDPR and AI Act, the U.S. AI Bill of Rights. There are also non-binding frameworks like Responsible AI, Fairness, Accountability, Transparency, and Ethics (FATE), etc. For the purpose of our inquiry, we consider them under compliance following the treatment of \textcite{saxena2026ethical}.}

\rev{\subsubsection{Privacy}
\textcite{saifullah2024privacy} investigated the effects of privacy-preserving techniques, specifically differential privacy and federated learning, on classical XAI methods for deep learning models. They argue that understanding the concurring objectives is crucial in improving XAI. This is pursued by a number of authors because explainability can inadvertently leak sensitive training data \parencite{shokri2021privacy}.  For instance, \textcite{abbasi2023} formulate a three-way trade-off among privacy, explainability, and data utility in machine-learning-based tabular data analysis, showing that privacy-preserving mechanisms may alter both predictive performance and the quality of explanations.}

\rev{\subsubsection{Fairness}
Taxonomy is proposed by  \textcite{ramachandranpillai2025fairxai} that aims to promote synergy between fairness and explainability. They specifically mention tradeoffs exist in AI systems  between procedural fairness and prediction accuracy, and fairness measures and interpretability.}

\rev{\textcite{stevens2020} constructed an algorithm that improved fairness while maintaining a high model performance and explainability on the dataset of Kiva, a loan lending platform for microfinance institutions and social impact businesses. In their study, one of the four models could have high fairness and performance, but post-hoc explanations (explainability) were not available.}

\rev{\textcite{saxena2026ethical} presented a lending architecture that achieves GDPR- and Basel III-oriented compliance alongside fairness and predictive performance by combining embedded regulatory checks with real-time explainability.}

\rev{\subsection{Cost}
\textcite{burgt2020explainable} discusses the potential cost of favoring simple, fully explainable models that only achieve 70\% accuracy over complex, less explainable models but with 99.99\% accuracy, particularly in critical tasks in banking like analyzing legal contracts. \textcite{thomas2026machine} show that neural networks have higher computational cost than gradient boosting machine when applied on credit risk assessments and in joint consideration with explainability, recommend the latter instead, but still recognize its utility for unstructured and datasets with \prf{a} high number of parameters.}

\rev{According to \textcite{dessain2023}, the cost of explainability is around 15 to 20 basis points in annual return on investment between best performing black boxes and inherently explainable models. This is affected by risk appetite increases of the bank in their credit scoring models. A solution proposed is incorporating expert judgment to not compromise financial performance. However, \textcite{YeoVanDerHeeverMaoEtAl2025} notes that even within surrogate models, there is a computational cost related to choosing a level of explainability, e.g., attention mechanisms are more efficient than RNN-based models.}

\rev{\subsection{Latency}
The demand for real-time XAI is also called \textit{in time explainability}, which is used in autonomous agents for self-driving cars on land or aerial and traffic control networks \parencite{Alzetta2020InTime}. Speed has also nevertheless, now becoming a crucial consideration in finance applications like algorithmic trading where there are high frequency trading (HFT) data and turnaround time between perceiving market signals and placing a trade affects the profitability and exposure to higher systemic risks \parencite{krishna2025latency}. In fact, \textcite{devi2024strategic} notes that the difference of microseconds in HFT can determine profit and loss, which quantum computing aims to solve by speeding up the analysis of market conditions.}

\rev{Towards this end, \textcite{assis2023performance_explainability} explores the balance between performance (accuracy and response time) and explainability in an intelligent environment. They found that transparent models (Decision Trees, Logistic Regression, K-Nearest Neighbors) offered faster response times and greater interpretability, making them suitable for applications like financial auditing where quick decision-making and clarity are critical. Meanwhile, opaque models (Convolutional Neural Networks (CNN), Support Vector Machines (SVM), Random Forest) generally achieved higher accuracy but incurred longer response times, especially under heavy workloads, demonstrating a trade-off with computational efficiency.}

\rev{\textcite{crook2023revisiting} also proposes an expansion of PET with a temporal budget as an additional dimension (PET+), but they assume the constancy of domain expertise and cost. The authors do acknowledge that this three-factor model do\prf{es} not capture all considerations in a model development. They were developed as an answer to the critique of \textcite{rudin2019stop} that resources should be allocated to the development of both highly performant and explainable models, especially in high stake domains. \textcite{rudin2019stop} notably called for the use of fully explainable models for ethical reasons and argued that inherently explainable models can compete with black boxes.}

\section{Methodology} \label{method}
To answer the research question, we relied on a qualitative design built around semi-structured interviews. This section outlines the interview setup, recruitment strategy, coding process, ethical safeguards, and the main limitations of the study.

\subsection{Exploratory Interviews}
We conducted twenty (20) semi-structured interviews with \prf{duration} ranging from 30--60 minutes between April and November 2025. Our sample size aligns with qualitative research literature indicating that thematic saturation of codes and meanings in similar studies is often achieved within 9--17 interviews \parencite{hennink2022}. The full interview guide was designed to support multiple papers and included broader questions on XAI expectations, explanation formats, and model interaction. However, only a subset of the guide was relevant to the present trade-off analysis. Accordingly, the analysis in this paper draws on three short segments of the \prf{interviews}:

\subsubsection{Role and Context}
\begin{itemize}
\item Can you briefly describe your role?
\item Do you currently interact with AI or machine learning (ML) systems in your work? In what way?
\item Are you mainly using outputs, developing/validating models, or overseeing projects that involve AI?
\end{itemize}

\subsubsection{Relevance of Explanation to Work Tasks}
\begin{itemize}
\item For the AI systems you work with, are explanations or justifications available in any form? What does that look like in practice?
\item Are there tasks where explanations matter more to you?
\end{itemize}

\subsubsection{\rev{Competing Factors} Ranking Task}
Participants were presented with five competing considerations for AI deployment and asked to rank them from most to least important:

\begin{enumerate}
\item Accuracy
\item Compliance
\item Cost
\item Ease of understanding
\item Speed
\end{enumerate}

\begin{itemize}
    \item Could you walk me through why you put them in this order?
    \item What made you place [factor X] above [factor Y]?
\end{itemize}

As seen above, participants were asked to rank them from most to least important. The items and their definitions were derived from the literature review (see section \ref{tensions}). Building on \textcite{habiba2025} who ask qualitatively about the accuracy–explainability trade-off for ML practitioners, we extend the focus to a broader set of operational factors and investigate how finance professionals themselves prioritize and justify these trade-offs. After ranking, participants were invited to explain their ordering and to describe concrete situations in which they would change or defend their priorities.

\subsection{Participant Recruitment}
\begin{figure}
    \centering
    \includegraphics[width=1.0\linewidth]{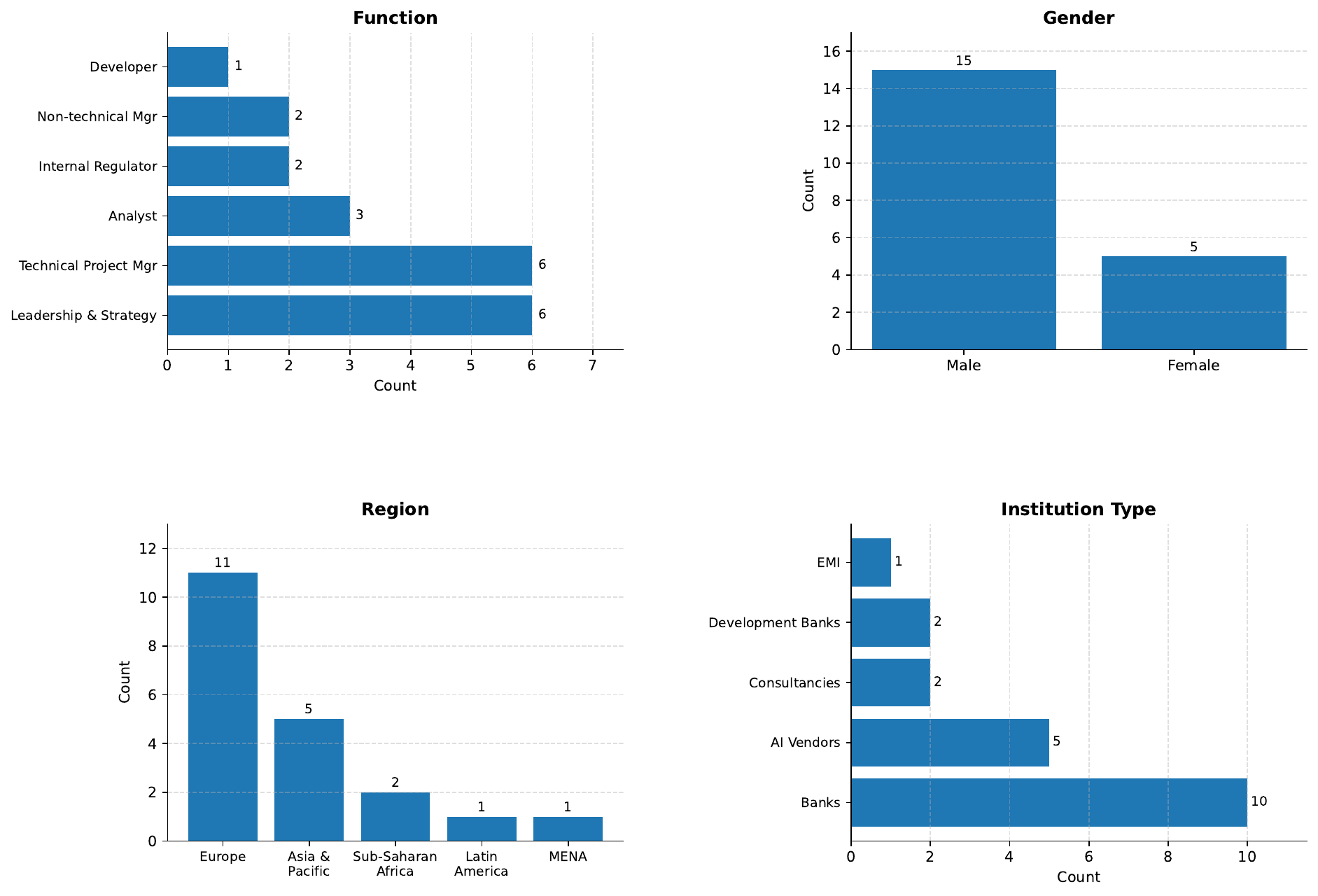}
    \caption{Summary of Interview Participants Demographics}
    \vspace{10pt}
    \small{Source: Authors' compilation}
    \label{fig:demography}
\end{figure}

The participants (Figure \ref{fig:demography}) \prf{\sout{The participants}} were selected through purposive and snowball sampling, beginning with initial contacts within the MSCA DIGITAL network and connections on various social media platforms. The recruitment process targeted professionals with demonstrable experience in the application or oversight of AI systems in financial contexts. Specifically, the inclusion criteria required that each participant:

\begin{itemize}
    \item is currently or has previously been employed in the financial sector, or in an auxiliary business primarily serving financial institutions (e.g., consulting or auditing firms, or technology vendors); and
    \item has directly interacted with, operated, or contributed to the development of an AI-based system or application for a period of at least six (6) consecutive months, either in a full-time or project-based capacity, i.e., as external consultants who work intensively, but may do or have done so for a limited period for a financial institution.
\end{itemize}

The first criterion focuses on the practical, real-world employment context within the sector, not academic background, to ensure participants have firsthand experience with the application of AI in the financial environment. We still included developers in the second \prf{criterion} because their role and workflow usually cover systematic logging of issues in software deployment that are reported by their users. These issues are indicative of the explainability needs of our target audience. The snowball approach allowed for a targeted but diverse pool of respondents from banks, AI vendors, electronic money issuers (EMI) and the consultancies providing auxiliary services to financial institutions. The participants’ backgrounds are summarized in Figure \ref{fig:demography} , referred to anonymously as P1-P20 in the results section (see Appendix \ref{appendix-a-participants} for the full list). 

\subsection{Data Analysis}
\rev{Qualitative analysis of the interview transcripts was conducted independently by two researchers to strengthen analytical rigor. The interviews were originally coded in full as part of a broader research project on explainability in financial AI. For the present article, two researchers independently coded the segments relating to the ranking task and discussion: one in Taguette, an open-source tool, and the other in a structured Word template, applying the five trade-off factors (accuracy, ease of understanding, compliance, cost, and speed) as organizing codes. After this independent round, the researchers compared their coding line by line, discussed any discrepancies, and quote groupings until they reached consensus on which supporting quotes are essential to answer the research question. In keeping with qualitative, interpretive approaches, consistency was ensured through this consensus-based reconciliation process rather than through calculating a formal intercoder reliability statistic.}

\rev{We also use descriptive statistics but only for the purpose of showing the weights of how each aspect was ranked first, second, etc. It allows us to illustrate how many of the participants value a certain \prf{aspect} in a succinct manner, without having to add all the relevant quotes to support the hierarchies. For this part, we use simple averages of the ranking and tallies.}

\subsection{Ethics Statement}
To ensure ethical conduct and participant privacy, all interviews were conducted with informed consent. The study’s purpose and participants’ rights were outlined in the information sheet, including their right to withdraw at any time. The study has also been reviewed and approved by the Ethics Committee for Research Involving Human Subjects in Biomedical Fields of the University of Naples Federico II, on September 25, 2025, with reference number PG/2025/0133225. 

\rev{\subsection{Limitations}}
\rev{This study is exploratory and qualitative in nature, and the ranking task was used mainly as an elicitation device rather than as a definitive measure of priorities. Since the factors were presented in a fixed set and order, the exercise may have shaped participants’ responses or crowded out considerations beyond the options provided, and the rankings should therefore be read as exploratory rather than conclusive. However, participants were also given space to explain, challenge, and qualify their rankings, allowing the analysis to focus on the reasoning behind their choices rather than on the ordinal positions alone.}

\rev{At this stage, the study also does not systematically include external regulators, even though regulation emerged as a central theme, so our account of regulatory expectations is based primarily on how these are understood and operationalized by professionals involved in the use, development, implementation, and internal oversight of AI systems. The sample is purposive, relatively small, and skewed toward European institutions, which limits broader generalization, while the reliance on self-reported experience means that practices cannot be independently verified and silent failures may remain unobserved. These limitations suggest caution in interpretation and indicate the value of future research that includes more diverse regional settings, more balanced participant profiles, and direct engagement with regulatory authorities.}

\rev{\section{Empirical Context: Use Cases and Participants } \label{context}
This section provides an overview of the AI use cases discussed in the interviews and the institutional and functional backgrounds of the participants who reflected on them. 
\subsection{Model Transparency in Applications}
\begin{figure}
    \centering
    \includegraphics[width=1\linewidth]{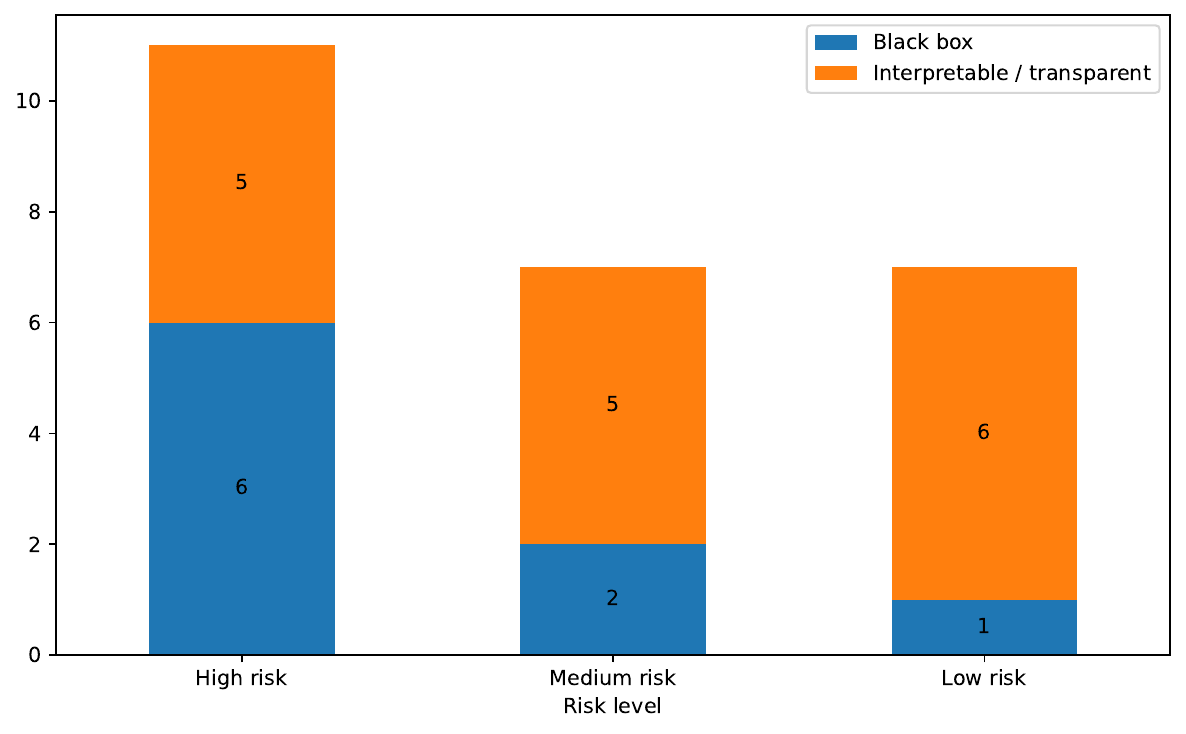}
    \caption{AI Use Cases by Risk Level and Model Transparency}
    \label{fig:usecases}
\end{figure}

Across the 20 interviewed professionals, 25 distinct AI use cases were identified, since several participants discussed more than one application in their organizational setting. When categorized by risk level (Figure \ref{fig:usecases}), 11 use cases fell into the high-risk category, 7 into the medium-risk category, and 7 into the low-risk category. High-risk cases were concentrated in credit scoring, lending, insurance pricing, and Anti-Money Laundering (AML) or fraud detection, where AI systems directly shaped financially consequential decisions and were therefore most exposed to regulatory scrutiny. Medium-risk cases were mainly found in corporate monitoring, research support, and advisory contexts, while low-risk cases clustered around internal productivity, customer support, policy checking, and coding assistance. Across these use cases, 9 applications were classified as black-box and 16 as interpretable or transparent, suggesting that even within a finance setting often associated with opacity, participants frequently described systems designed to remain traceable, auditable, or human-readable in practice.}

\rev{\subsection{Business Users}
Following audience-based role distinctions in explainable AI research, we focus here primarily on finance professionals as one of the main groups who receive, use, or rely on explanations, alongside regulators and the public \parencite{tomsett2018, arrieta2020, jiang2022who, Wilson2025CFA}. 

At the role level, they included owners or leadership and strategy professionals (P1, P3, P7, P11, P14, P19), product managers or technical project managers responsible for coordinating deployment between business needs and system design (P2, P5, P8, P12, P15, P17), operators or non-technical managers and analysts who worked directly with AI-supported outputs in practice (P4, P6, P9, P13, P16), creator-developers who built or adapted the system itself (P20), and internal regulators, that is, internal risk, compliance, or control functions concerned with model oversight and organizational accountability (P10, P18).}

\rev{While technically oriented participants, including creator-developers and product managers, formed a substantial part of the sample, their inclusion was justified by the study's focus on real-world implementation, where such actors routinely mediate between end users, organizational requirements, and system design.}

\rev{\subsection{Organizational landscape}
Of the 20 participants, 10 worked in banks, 2 in development banks, 1 in an electronic money issuer, 5 in AI vendors, and 2 in consultancies providing auxiliary services to financial institutions. Rather than treating all non-bank participants as a single residual category, it is important to note that the AI vendors in our sample were not generic technology firms, but organizations whose products, implementation work, or compliance-related services were closely tied to financial-sector use cases. This follows the recruitment criteria of the study, which explicitly included professionals employed either in the financial sector itself or in auxiliary businesses primarily serving financial institutions, provided that they had direct experience with AI systems in financial contexts.}

\rev{We would like to note, however, that P18 came from a values-based bank and is therefore best treated within the banking category, albeit as a philosophically distinctive case. P19, by contrast, came from a bespoke technology organization with an NGO orientation, but was included because the interview focused on a finance-specific AI use case consistent with the study’s recruitment criteria for auxiliary organizations serving financial institutions. The sample therefore includes not only incumbent financial institutions but also a smaller set of finance-adjacent technology and service \prf{providers} that participate in the design, deployment, or oversight of AI in practice.}

\section{ \rev{Competing Factors in} AI Deployment as Perceived by Finance Professionals} \label{tradeoff_results}
The findings begin with the overall ranking pattern across the five factors and then move to the reasoning behind those choices. From there, the section examines how participants justified priorities, where they differed, and what other concerns emerged during the interviews.
\begin{table}[htbp]
\centering
\caption{Competing Factors Rankings Summary of Central Tendencies and Extremes}
\label{tab:ranksummary}
\small
\setlength{\tabcolsep}{3pt}
\begin{tabular}{p{3.2cm}rrrrrr}
\toprule
Factor & Mean rank & Median rank & Rank 1 & Top 2 & Bottom 2 & Rank 5 \\
\midrule
Accuracy & 2.35 & 2 & 6 & 12 & 4 & 0 \\
Compliance & \prf{2.63}& 2 & 6 & 11 & 6 & 3 \\
Ease of Understanding & 2.65 & 2.75 & 6 & 9 & 7 & 2 \\
Cost & \prf{3.48}& 3.25 & 1 & 5 & 9 & 6 \\
Speed & \prf{3.90}& 4 & 0 & 2 & 13 & 4 \\
\bottomrule
\end{tabular}

\vspace{4pt}
\begin{minipage}{0.95\textwidth}
\footnotesize
Note: Rankings are summarized descriptively only. When participants reported ties, tied items were assigned the average of the rank positions they jointly occupied (e.g., a two-way tie for third/fourth received rank 3.5). Mean and median ranks are based on these averaged values. Top-2 and Bottom-2 counts include tied responses only when the averaged rank fell within those ranges.
\end{minipage}
\end{table}
\rev{Table \ref{tab:ranksummary} summarizes the rankings of the 20 interview participants across the five aspects. Accuracy was the most consistently prioritized factor, with the lowest mean rank, the highest concentration in the Top 2, and no last-place rankings, indicating that participants largely treated it as a baseline requirement for trustworthy AI use. Compliance and ease of understanding also ranked relatively highly, although ease of understanding showed greater dispersion across participants, while cost and especially speed were more often placed in the lower part of the ranking.}

\subsection{Accuracy}
\rev{Accuracy emerged as the most consistently prioritized aspect, combining the lowest mean rank \prf{2.35}, a strong Top-2 concentration \prf{12}, and no Rank-5 placements, which suggests that respondents largely viewed it as a non-negotiable prerequisite.} The rationale for accuracy being a top priority is directly tied to the functional purpose and credibility of the AI system. The respondents’ perspective is grounded in the understanding that an AI system is worthless if it is not correct. 

When an AI fails to provide correct, factually grounded information, or when it presents incorrect information with unearned confidence, it fundamentally erodes not only its utility, but also the user's trust. As P5 puts it, \textit{``if the answer is not accurate... he or she [the client] would never use the tool again... trust is very much crucial. And it’s the credibility we have as a research house [of the bank].”} 

Participants conceptualized accuracy differently depending on whether they were working with classical machine learning models or generative AI systems. In classical ML contexts (e.g., credit scoring models or \rev{Anti-Money Laundering (AML)} transaction monitoring), accuracy was linked to performance metrics, i.e., reducing false positives and false negatives, stabilizing model behavior over time, and ensuring reliability across monitoring cycles. P8 described quarterly monitoring using R-squared (coefficient of determination) and absolute error metrics to identify issues such as a \textit{``dip in the input data” }that could undermine outputs. 

By contrast, in generative AI systems (e.g., internal Q\&A assistants with LLM-assisted responses or RAG-based document search tools), accuracy was defined as factual faithfulness to source documents rather than statistical performance \rev{or fidelity of the surrogate models with the black box systems, as in ML applications}. P2 described accuracy as ensuring that the answer \textit{``really, really comes from the source, not that it was somehow generated”}, supported by automated open-source frameworks that check whether outputs match original materials. 

Developers also cautioned that LLMs \rev{as an explanatory layer} can be \textit{``overconfident with the answer,” }resulting in a contradictory effect:  the human-readable language of the interface helps users explain the model (a feature P17 appreciated), yet, as P16 warned, the explanation may ultimately bear little relation to the actual input. \rev{Such \prf{issues}, popularly known as \textit{hallucination\prf{s}} have} been found pervasive in LLMs \parencite{jacovi2020} \rev{and an open research question \prf{in} this regard in finance is the existence of a universally applicable index to objectively assess their capabilities and explainability in financial scenarios \parencite{liu2024research}.} 

Accuracy was prioritized whenever errors risked undermining explainability, confusing users, or exposing the institution to regulatory or reputational consequences (see Table \ref{tab:accuracy_cases}). 

\begin{table}[htbp] 
\centering
\caption{Cases for Prioritization of Accuracy \label{tab:accuracy_cases}}
\vspace{5pt}
\small Source: Authors' transcripts
\begin{tabular}{p{2cm} p{4.5cm} p{5cm}}
\toprule
\textbf{Factor} & \textbf{Relationship}& \textbf{Supporting Quotes} \\
\midrule
Accuracy over Ease of Understanding& 
Accuracy of the explanation is necessary to validate the underlying model's performance; inaccuracy leads to misleading outputs. & 
``if the explanation isn't accurate, then it's just a double issue… it creates some kind of strange mix of false positives and negatives… I don't know what even was happening'' (P2). \\
\midrule
Accuracy over Compliance& 
Compliance is endangered if the system is inaccurate, leading to severe institutional consequences. & 
``If the AI produces `errors or false data,' the financial institution is `liable' to the regulator, resulting in penalty and reputational damage'' (P14). \\

 &  & 
Factualness is essential to ``validate whether the internal guidelines is [are] meeting external guidance requirements (sic)'' (P2). \\
\bottomrule
\end{tabular}
\end{table}

\rev{Meanwhile, for the times accuracy was a secondary concern, the reasons are as follows: the models are already \textit{super stable and predictable} (P8) or \textit{because there is a human in the loop} (HITL) to verify outputs (P7). The other instances are because compliance, ease of understanding or cost were considered higher in importance.}

\subsection{Compliance}
\rev{Compliance was frequently placed near the top of the ranking, with \prf{11} Top-2 placements and a mean rank of \prf{2.63}, suggesting that many participants did not treat it as a factor to be taken lightly, but as a standing condition of legitimate AI use. In the interviews, this was reflected in the way compliance was discussed not only as adherence to external law, but also as the internal translation of those legal expectations into organizational controls, approval processes, and risk thresholds.}

Finance is notorious for being highly regulated. P11 puts it as \textit{``regulated to death."}  \rev{This relationship necessitates \prf{a}  compliance function as the internal agent responsible for operationalizing a bank's  policy to manage risk \parencite{edwards2004compliance}. It involves reacting to requirements formulated by entities external to the financial institution. The internal and external aspect of the law is apparent when participants described their ranking for compliance. For example, P17 mentioned the EU AI Act when asked for other factors that might be relevant to the tradeoffs. On the one hand, there are external regulations set by institutions like the EU, European Banking Authority and counterpart central banks. On the other, there are the equivalent guidelines internal to the financial institution that translate them into actionable rules.}

\rev{Although ease of understanding is a separate aspect of AI deployment in our questions, compliance is tied to explainability for some participants.} Dubbed by a participant (P17) as \textit{``big compliance machines,”} banks face multiple regulatory mandates, especially in the Eurozone. For example, as reported by P7 and P8, the European Central Bank (ECB) and European Banking Authority (EBA) do not allow black box models (such as neural networks or complex boosting techniques) for use in decision-making processes because they imply \textit{``unpredictability and the risk involved.”} \sout{Although the EU AI Act do not prohibit black boxes in finance  completely and rather it use a risk-base system,  there are some AI practices that are prohibited outright under Article 5 \parencite{EUAIAct_AnnexIII_Web}.  These include creditworthiness assessment and credit scoring of natural persons, are classified as high-risk and are therefore allowed subject to strict requirements rather than banned. The responses of the participants illustrate the more conservative response of internal teams to regulations.}

\prf{Although the EU AI Act does not prohibit black-box models in finance as a general category, it applies a risk-based framework. Certain AI practices are prohibited, while AI systems used to assess the creditworthiness or credit score of natural persons are classified as high-risk and are permitted subject to specific requirements. \cite{EUAIAct_AnnexIII_Web}}

In Table \ref{tab:compliance_cases}, we present cases when compliance is preferred over accuracy and ease of understanding. 
\begin{table}[htbp]
\centering
\caption{Cases for Prioritization of Compliance}
\label{tab:compliance_cases}
\vspace{5pt}
{\small Source: Authors' transcripts}
\begin{tabularx}{\textwidth}{
  >{\raggedright\arraybackslash}p{2cm}
  >{\raggedright\arraybackslash}X
  >{\raggedright\arraybackslash}p{2.5cm}
  >{\raggedright\arraybackslash}X}
\toprule
\textbf{Factor} & \textbf{Relationship} & \textbf{Compliance Dimension} & \textbf{Supporting Quotes} \\
\midrule
Compliance over Others &
Compliance requires satisfying auditors and regulators. & External &
``The explainability [\ldots] for the regulator mostly.'' (P10) \\
\midrule
Compliance over Ease of Understanding &
Compliance is a non-negotiable prerequisite. & Internal &
``I just put alignment with regulations first [\ldots] So it is a given anyhow.'' (P2) \\
\bottomrule
 &  & Internal &
``Without having compliant processes [\ldots] So it's quite on top.'' (P5) \\
\end{tabularx}
\end{table}

Notably, even for those who rank compliance low, this does not imply that compliance is unimportant. Rather, they consider it a ``given” (P2) at the company level, making it a ``hygiene factor,” a basic precondition that must be satisfied before other factors (like accuracy or speed) are even considered (P2, P11, P6, P17). By using a \textit{``compliance by design”} (P13) approach, so that\textit{ ``whatever regulations that come… you can’t run far (from complying) because your system is already doing the right thing.”} P1, on the other hand, ranked compliance last because the AI prototype tool was already limiting the scope of information provided internally. 

\subsection{Ease of understanding}
\rev{Ease of understanding ranked relatively high but showed greater dispersion than accuracy, with a mean rank of \prf{2.65}, nine Top-2 placements, and seven Bottom-2 placements. Its broader spread across the ranking distribution indicates that participants did not value it uniformly, instead prioritizing it most when adoption, communication, or cross-functional usability were at stake.}  

P6 \prf{overrode} accuracy with ease of understanding noting that, \textit{``no matter how accurate or how fast or how compliant it is, if no one will understand it, then no one will consume it."}  P9 did so as well because it enables financial investigation units, financial institutions and regulators to \textit{``be on the same page quite quickly."} A unique insight from P18 who is a CTO in a values-based bank is that this aspect is \prf{a} priority given the deliberate philosophy of the institution that prioritizes the human being. The goal is to ensure employees fully comprehend the tool before analyzing its accuracy, safeguarding their critical thinking skills.

\begin{table}[htbp]
\centering
\caption{Cases for Prioritization of Ease of Understanding \label{tab:ease_cases}}
\vspace{5pt}
\small Source: Authors' transcripts
\begin{tabular}{p{2cm} p{4.5cm} p{5cm}}
\toprule
\textbf{Factor} & \textbf{Relationship} & \textbf{Supporting Quotes} \\
\midrule
Ease over Others& 
If an explanation cannot be understood, its accuracy is irrelevant. & 
``if I cannot communicate what I'm doing [...] no one will consume it.'' (P6) \\
\midrule
Ease over Accuracy& 
Participants have a \rev{perceived} willingness to accept some loss of accuracy.& 
``[Talking to the bank's] compliance, they say, if I have an extremely performant model, but which is a bit harder to explain towards a regulator versus a medium performing algorithm, which is perfectly explainable [...] always go for the second one...you go for explainability and you really need to balance it off towards performance. Performance is not the first measure to select tooling.'' (P10)\\
\bottomrule
\end{tabular}
\end{table}
In contrast, ease of understanding is sacrificed when dealing with already highly expert audiences, strict regulatory environments, or tasks where errors carry severe reputational risks. P5 ranked this last because their target audience (portfolio managers and institutional clients)\textit{``has quite profound knowledge already, and it usually doesn't really matter so much if the answer is more complex." } Similarly, P7 also put it as least important because their users are seasoned financial experts and small-, medium- enterprises to whom the complex signals are already very familiar. This supports prior findings by \textcite{bayer2022role} where experts use explanations to verify their previous knowledge or experience. 

\subsection{Cost}
\rev{Cost carried a moderate but secondary weight in participants’ rankings, with a mean rank of \prf{3.48}, only one first-place ranking, five Top-2 placements, and \prf{nine} Bottom-2 placements, indicating that it was usually treated less as a primary objective than as a feasibility constraint on deployment. Direct cases of cost being prioritized over other factors are limited, but some participants clearly treated it as a decisive feasibility condition, especially when AI investments had to be justified in terms of manpower savings, usage costs, or overall business benefit.}

As P1 noted, budgetary constraints are strict and dictate feasibility of AI systems as a whole: \textit{``if we don’t have the budget for that, it will simply not fly.”} Even though other participants were not exactly privy to the financial sheets of the company, most still emphasized the necessity of a rigorous cost-benefit analysis (P5, P7, P13, P16) which changes over time. This aligns with the concept of ``hidden technical debt” in ML systems, the accumulation of unseen maintenance and integration burdens behind apparently simple models \parencite{sculley2015}, suggesting that while rapid deployment offers immediate gratification, it often obscures long-term maintenance challenges and infrastructure costs. As P5 argued, high investment \textit{``must be traded with the benefit you get out of it.”} For example, in the case described by P6, the post-implementation of black-box models for alternative credit scoring resulted in a 40\% expansion of the bank’s credit card base, bringing in approximately \$5 million in transactions. 

On the other hand, the weight given to the cost \rev{of explainability, in particular,} is heavily contingent on the scale and complexity of AI models deployed. P4 qualified that cost becomes critical when dealing with computationally intensive tasks, such as counterfactual explanations which require \textit{``a very expensive search if it’s a big model.”} This distinction was reinforced by participants from the banking sector. P2 noted that their smaller scale of AI application meant they did not incur \textit{``major costs from running explanations,” }whereas P7, operating at a larger volume, emphasized the need to be \textit{``cost efficient because the cost/call centers really look at how many API calls they do.”} 

\rev{Interestingly, P10 defines performance as reduction of operational cost tied to a reduction of false positives, effectiveness and finding new types of money laundering. This shows that accuracy (normally expressed through metrics like false positives or recall) are also considered together with cost since resources are spent on cases that otherwise would not be needed. When the effects of low accuracy are translated to business impact, cost is then naturally embedded within the aspect of performance.}

\subsection{Speed}
\rev{Speed carried the weakest overall weight in the ranking exercise, with the highest mean rank at 3.90, no first-place rankings, only two Top-2 placements, and \prf{13} Bottom-2 placements. This \sout{can be due to the covered} \prf{may be because the} banking and supervisory processes operate in slow-moving cycles of days or months—a select few professionals (such as P2, P7, P8, P14) elevated it to their 2nd most important priority (see Table \ref{tab:speed_cases}).  This relatively low ranking of speed should be interpreted cautiously, because the literature treats latency as far more consequential in high-frequency trading and other real-time financial applications, whereas such use cases were not represented in our interview sample.}

\begin{table}[htbp] 
\centering
\caption{Cases for Prioritization of Speed \label{tab:speed_cases}}
\vspace{5pt}
\small Source: Authors' transcripts
\begin{tabular}{p{2cm} p{4.5cm} p{5cm}}
\toprule
\textbf{Factor} & \textbf{Relationship}& \textbf{Supporting Quotes} \\
\midrule
Speed over Ease of Understanding& 
Responsiveness matters more for perceived usefulness.& 
``giving something faster gives a better feeling to the users than how easy the answer is" (P2)\\
\end{tabular}
\end{table}

A distinct pattern emerged regarding the necessity of speed in both AI model answers and its explanations, where applicable, delineated by the stage of the AI lifecycle. Technical personnel involved in the development phase generally viewed computational speed as a secondary concern. As P2 explained, workflows that are not yet tied to real-time transactional systems allow for significant latency: \textit{``I don’t think we are worried that it comes with[in] 10 seconds... or maybe even 10 minutes? We would press a button in the evening and come in the morning.”} 

However, this tolerance disappears once the system transitions to operational use. Participants acknowledged that \textit{``for a user interface, it’s important to give everything fast,}” recognizing that once the system is deployed for operational use, the speed of responsiveness becomes an important user-facing requirement. For front office professionals like P3 and P14, speed is strictly dictated by key performance indicators (KPIs), or metrics that evaluate the quality of service delivered in a specific banking role. In customer support, this can be the time it takes for an employee to resolve a complaint. 

P14 noted that financial crime investigators have hard targets to \textit{``clear the cases in one to three days,”} while P1 highlighted that customer service departments operate under strict timeframes for interaction. Consequently, latency becomes an adoption barrier; P1 warned that if a model \textit{``would take one hour, I don’t think anyone would use it. They would rather type.”}  This observation aligns with \textcite{kruse2019}, who argue that if an AI tool (such as a chatbot) lacks speed, its ``added value is questionable.” 

\rev{\subsection{Audience-dependent tradeoffs}}
\rev{The interviews further suggest that the prioritization of the presented competing objectives are role-dependent even within the same institution. The bank-level comparisons show that employees from the same bank did not converge on a single ordering of priorities; rather, their rankings varied systematically with role, reflecting different responsibilities in deployment, validation, operational use, and oversight.}

\begin{figure}
    \centering
    \includegraphics[width=0.75\linewidth]{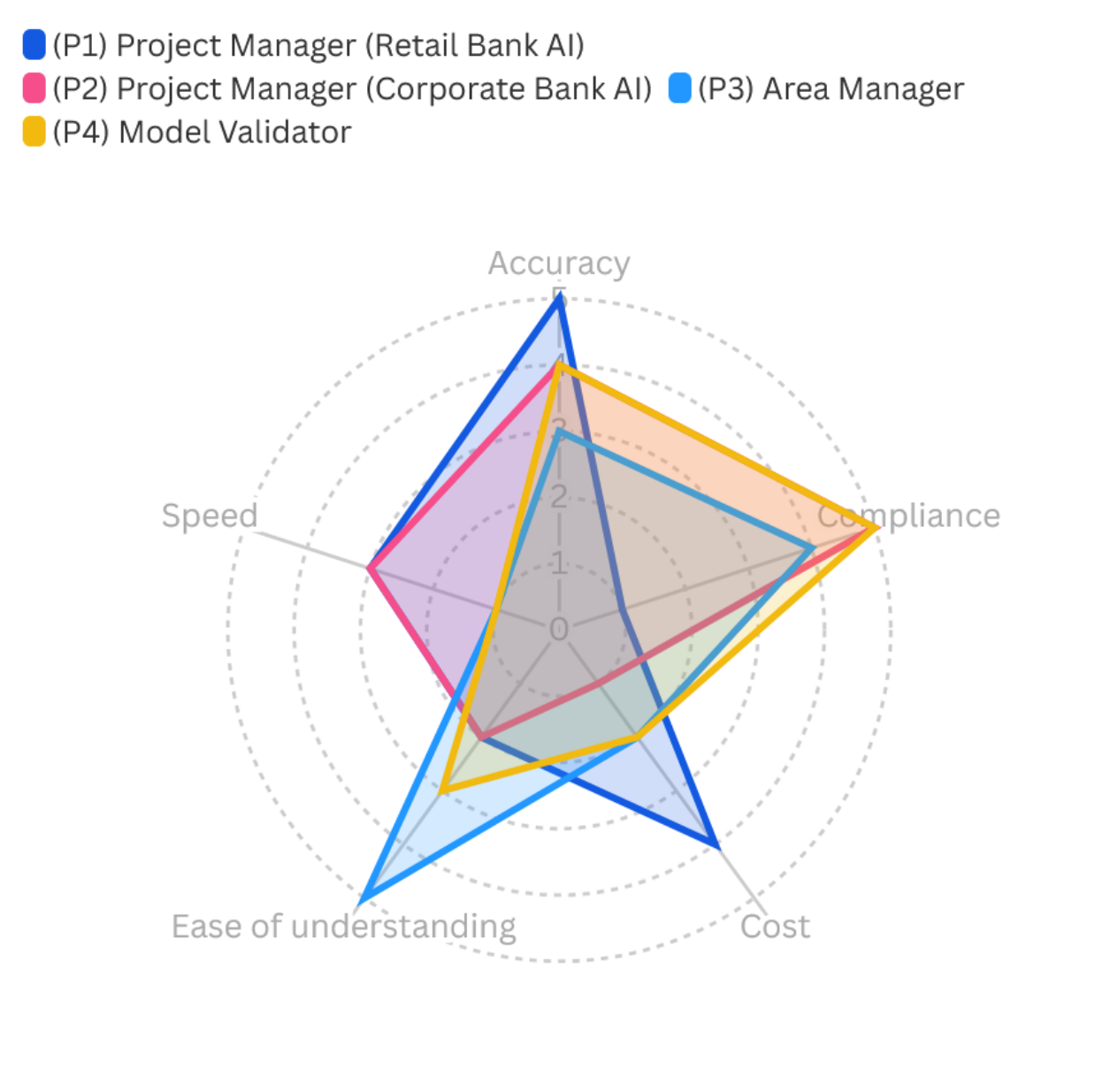}
    \caption{Tradeoff Rankings by Bank A participants}
    \label{fig:bank_a}
\end{figure}

\rev{For example, in Bank A, P1-P4's rankings differ drastically because they occupy different roles and interact with completely different AI use cases (Figure \ref{fig:bank_a}). P1 prioritizes Accuracy and Cost to get a functional product launched; P2 prioritizes Compliance and Accuracy to avoid fines while building internal RAG tools; P3 prioritizes Ease of Understanding to debug customer interactions; and P4 prioritizes Compliance above all else because ECB audits act as an absolute barrier to entry for credit risk modeling. Surprisingly, for P4 who is a technical personnel, Accuracy is ranked at the Bottom-2 with Speed but this is because, they are forced to use simple, stable linear or logistic regression models where accuracy is considered solved with a predictable baseline.}

\begin{figure}
    \centering
    \includegraphics[width=0.75\linewidth]{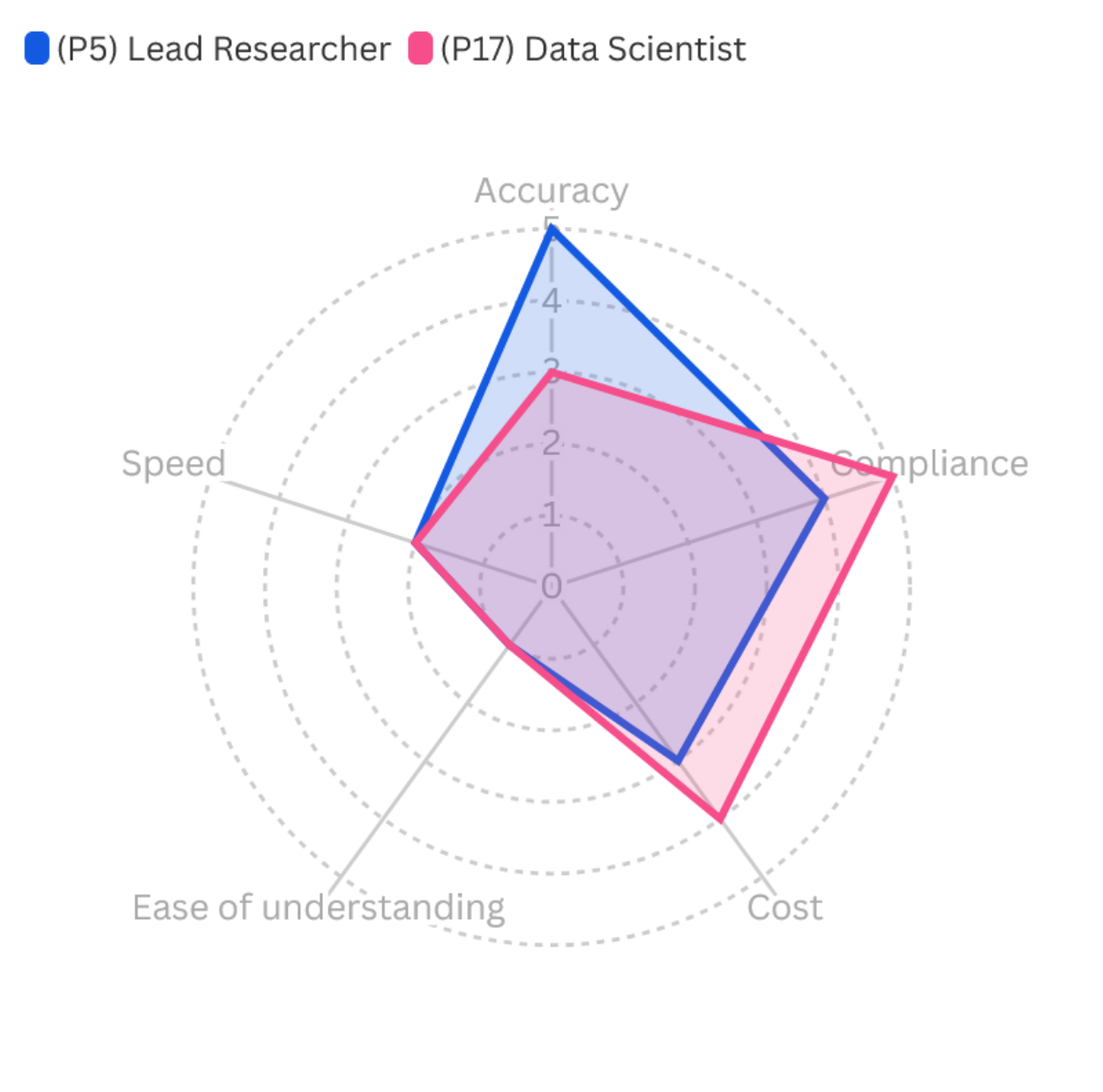}
    \caption{Tradeoff Rankings by Bank B participants}
    \label{fig:bank_b}
\end{figure}

\rev{For Bank B (Figure \ref{fig:bank_b}), P5 and P17's rankings differ because their AI tools serve different purposes. Both treat compliance and cost as important constraints, but P5 puts accuracy first because errors would damage client trust in externally facing research, whereas P17 prioritizes compliance because the internal monitoring tool is judged mainly on governance and feasibility. Ease of understanding is low for both, but for different reasons: P5’s users are expert clients, while P17’s users mainly want fast, relevant outputs.}

\begin{figure}
    \centering
    \includegraphics[width=0.75\linewidth]{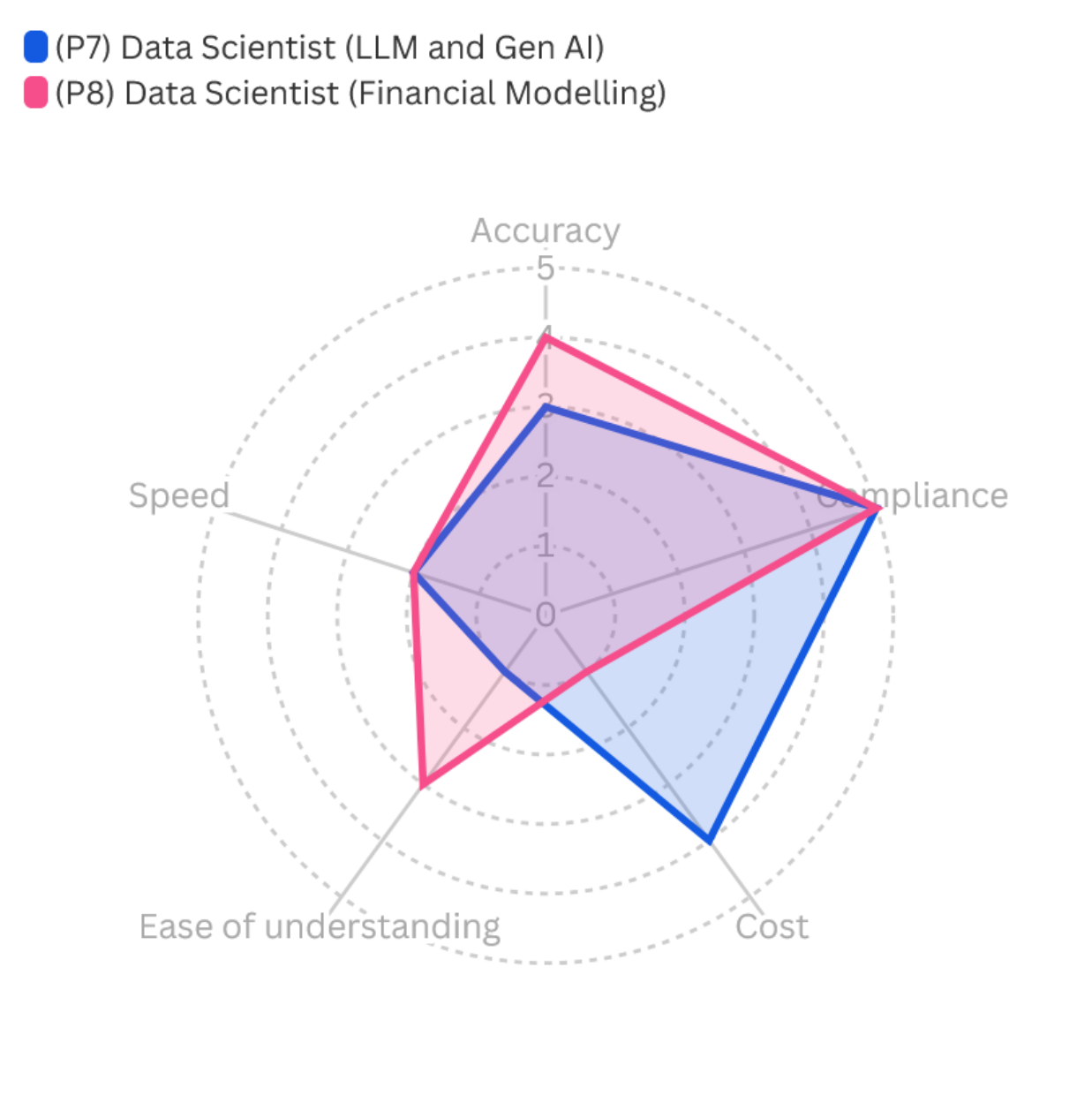}
    \caption{Tradeoff Rankings by Bank C Participants}
    \label{fig:bank_c}
\end{figure}

\rev{Despite operating in the same banking environment, priorities of participants from Bank C (Figure \ref{fig:bank_c}) diverge sharply based on scale and human involvement, while uniting entirely on regulatory boundaries. P7 must prioritize Cost because their AI application processes hundreds of thousands of documents, generating massive API overhead. P8 works on contained, internal mathematical models where cost is a negligible factor. Both of them do not put accuracy as the first because of the presence of HITL and because they use explainable models anyway, respectively for P7 and P8. Further, they rank Compliance as their absolute number one priority.}

\rev{\subsection{Other reported factors}}
\rev{Below, we summarize the other factors mentioned by the participants when given the time to report any other considerations beyond the five presented to them.}
\rev{\begin{enumerate}
    \item \textbf{Data Privacy, Security, and Confidentiality:} Data protection is frequently cited as a massive hurdle, sometimes overshadowing the AI's capabilities. For P19, data protection and the secure handling of Personally Identifiable Information (PII) is considered the biggest barrier to adoption for NGOs who use their technology for cash transfers with marginalized populations. Within corporate consulting (P12), the primary fear is leaking internal data to off-the-shelf AI systems, leading companies to strictly restrict AI usage. As a response, P8 from Bank C, says this pushes them to create their own models. In our study's treatment, privacy could be understood by laws like GDPR, but this shows that it may be useful to have it as a separate aspect for tradeoff following \textcite{saifullah2024privacy, shokri2021privacy, abbasi2023}.
\item \textbf{Data Quality and Semantic Alignment:} Before an AI can even be accurate or explainable, P11 reports that teams must first resolve semantic conflicts. If the humans cannot agree on the definitions of the variables the model will ingest, the AI will generate plausible but entirely meaningless results. Furthermore, when AI models generate unintuitive or seemingly incorrect financial relationships, data scientists often find that the root cause is simply inaccurate or bad data recorded in the bank's columns, rather than a flaw in the AI's logic. This aligns with the finding of  \textcite{nakashima2024users} that for developers of AI systems in retail banking, explanations were irrelevant for their trust in black box systems, and rather data quality and methodological choice factors were more determinant. 
    \item \textbf{Deterioration of Critical Thinking Skills:} P18, like most others, rank\prf{ed} speed at the bottom, but coming from a values-based bank, they raised a profound reason on why it would be the case. Their company is concerned regarding how AI affects the human worker. As such, they deliberately are slow in adoption since there is a discussion on whether relying on AI copilots will deteriorate employees' fundamental skills, such as writing, analyzing, and critical thinking over time. Researchers have termed this as ``cognitive surrender" \parencite{orland2026cognitive_surrender}, a dynamic that in finance can lower scrutiny in high-stakes decisions with potentially serious human, regulatory, and reputational consequences. This way, it can be seen that accuracy does not only rely on the AI but also on the humans in operation.
\end{enumerate}}

\section{Conclusion} \label{conclusion}
\rev{The exploratory study shows that finance professionals do not approach AI deployment as a simple trade-off between accuracy and explainability, but consider them along with many factors. In the ranking exercise where we presented five factors, accuracy is \prf{generally} prioritized, while explainability (operationalized through ease of understanding) is only second to compliance. \textit{Accuracy} received the highest mean rank at 2.35, followed by \textit{Compliance} at 2.625 and \textit{Ease of Understanding} (how we operationalize explainability for non-technical users) at 2.65, while \textit{Cost} at 3.475 and especially \textit{Speed} at 3.9 ranked lower overall.}

\rev{With the supporting quotes from the participants, this suggests that accuracy and compliance function largely as baseline conditions of acceptable AI use, whereas ease of understanding becomes most important when adoption, communication, and cross-functional accountability are at stake. Cost and speed, by contrast, were usually treated less as primary objectives than as practical constraints on what can be implemented and sustained in organizational settings. The findings also show that these priorities are role- and context-dependent rather than fixed across institutions, since participants in the same organization often ranked the factors differently depending on whether they were building, validating, governing, or directly using the system.} 

\rev{Future studies could examine how competing objectives differ across banking use cases with distinct temporal and regulatory demands, for example comparing high-frequency trading, credit scoring, AML investigation, and customer-service copilots, where the relative importance of speed, compliance, and ease of understanding is likely to vary substantially. Another valuable direction would be to study real deployment settings through observation, audit documents, or case tracing, since self-reported interviews may miss silent failures, abandoned projects, and unspoken compliance tensions.}

\section*{Funding}
This project has received funding from the Marie Skłodowska-Curie Actions under the European Union’s Horizon Europe research and innovation program for the Industrial Doctoral Network on Digital Finance, acronym DIGITAL, Project No. 101119635.

\section*{Acknowledgments}
We would like to thank the anonymous reviewer from this special issue for their comments on the paper.

\printbibliography

\section*{Appendix}
\addcontentsline{toc}{section}{Appendix}
\subsection*{Background of Interview Participants}
\label{appendix-a-participants}

\begin{longtable}{|c|c|c|c|}
\hline
\textbf{\#} & \textbf{Gender} & \textbf{Type of Institution} & \textbf{Function} \\
\hline
\endhead
P1 & F & Bank A & Technical Project Manager \\
\hline
P2 & M & Bank A & Technical Project Manager \\
\hline
P3 & F & Bank A & Non-technical Manager \\
\hline
P4 & M & Bank A & Developer \\
\hline
P5 & M & Bank B & Leadership and Strategy \\
\hline
P6 & M & Bank D & Developer / Technical Project Manager \\
\hline
P7 & M & Bank C & Technical Project Manager \\
\hline
P8 & M & Bank C & Technical Project Manager \\
\hline
P9 & M & AI Vendor A & Leadership and Strategy \\
\hline
P10 & M & AI Vendor B & Non-technical Manager \\
\hline
P11 & F & AI Vendor C & Technical Project Manager / Internal Regulator \\
\hline
P12 & M & Consultancy A & Analyst \\
\hline
P13 & F & Consultancy B & Leadership and Strategy \\
\hline
P14 & F & Electronic Money Issuer A & Analyst / Internal Regulator \\
\hline
P15 & M & Development Bank A & Analyst \\
\hline
P16 & M & AI Vendor D & Leadership and Strategy \\
\hline
P17 & M & Bank B & Technical Project Manager \\
\hline
P18 & M & Bank E & Leadership and Strategy \\
\hline
P19 & M & AI Vendor E & Leadership and Strategy \\
\hline
P20 & M & Development Bank B & Analyst \\
\hline
\caption{Interview participant demographics and background information.}
\end{longtable}

\textit{Note:}  To avoid double counting, each participant was assigned to one functional category only, based on their highest or most overarching role within their organization. When participants held hybrid or overlapping responsibilities (e.g., developer and project manager, or manager and internal regulator), we classified them according to the function that best captured their seniority, decision-making authority, or regulatory relevance. For example, P6 was coded as a Technical Project Manager rather than a developer, and P11 was coded as an Internal Regulator rather than a project manager. 

\end{document}